\documentclass[prl,aps,showpacs,twocolumn,superscriptaddress]{revtex4}
\usepackage{graphicx}

\begin{document}

\preprint{APS/123-QED}

\title{Defect-induced perturbations of atomic monolayers on solid surfaces}

\author{H. Schiessel}
\affiliation{%
Max-Planck-Institut f\"{u}r Polymerforschung, Theory Group, POBox
3148, D 55021 Mainz, Germany
}%

\author{G. Oshanin}
\affiliation{Laboratoire de Physique Th\'{e}orique des Liquides,
Universit\'{e} Paris VI, 4 Place Jussieu, 75252 Paris Cedex 05,
France
}%

\author{A. M. Cazabat}
\affiliation{Laboratoire de Physique de la Mati{\`e}re
Condens{\'e}e, Coll{\`e}ge de France, 11 Place M. Berthelot, 75252
Paris Cedex 05, France
}%

\author{M. Moreau}
\affiliation{Laboratoire de Physique Th\'{e}orique des Liquides,
Universit\'{e} Paris VI, 4 Place Jussieu, 75252 Paris Cedex 05,
France
}%

\date{\today}

\begin{abstract}
We study long-range morphological changes in atomic monolayers on
solid substrates induced by different types of defects; {\it
e.g.}, by monoatomic steps in the surface, or by the tip of an
atomic force microscope (AFM), placed at some distance above the
substrate. Representing the monolayer in terms of a suitably
extended Frenkel-Kontorova-type model, we calculate the
defect-induced density profiles for several possible geometries.
In case of an AFM tip, we also determine the extra force exerted
on the tip due to the tip-induced de-homogenization of the
monolayer.
\end{abstract}

\maketitle

Equilibrium properties of monolayers adsorbed on ideal,
defect-free solid surfaces are by now reasonably well understood
through a series of experimental and theoretical
works~\cite{adamson}. However, most of naturally encountered
surfaces or surfaces involved in technological processes can not
be considered as ideal and do contain different types of defects,
such as, {\it e.g.} chemisorbed or adsorbed species or surface
steps. Experimentally, it has been well-known that such defects
may have a profound effect both, on the adsorption kinetics and on
the equilibrium morphology of the resulting layers. In particular,
point defects often constitute nucleation sites for the adsorbates
and serve as seeds for island formation~\cite{krug}. On the other
hand, in the presence of a monoatomic surface step the adatoms
 on the lower terrace are
generally attracted towards the
step, which
causes their redistribution within the layer, as
observed, for instance, via intensity
oscillations of thermal He scattering at grazing incidence in the
form of the discrete row growth of Xe on stepped
substrates~\cite{Marsico97,Pouthier97}. Theoretically,  the impact
of the surface steps on the adatom distribution
was studied within the framework of two-dimensional (2D)
lattice-gas-type models~\cite{Albano87,Merikoski94}.
These models have been analysed
numerically and have revealed inhomogeneous
density profiles with an enhanced density
close to the lower step edges. To the best of our knowledge,
however, the analytical solution of the problem is still lacking.

On the other hand, probing of the monolayer properties by
different experimental techniques, such as, e.g. the STM or AFM
measurements, may itself incur morphological changes into the
adlayer. The interaction of the adatoms with the AFM tip might
cause their displacement from the adsorption sites. Such
deformations have been predicted for solid surfaces
themselves~\cite{mayer} and were indeed observed in MD
 simulations~\cite{con}. The adatoms are, or course,
even more vulnerable to the presence of the AFM tip since they
are not so strongly connected as the atoms of the
solid.
Indeed, it has been demonstrated that the SFM tip
can be used to "drag" single atoms or
molecules on metal
surfaces~\cite{li96,bartels97,kurpick99,rahman}.
Moreover, it has been observed in recent
experiments~\cite{Bardon98}
that the apparent thickness of the prewetting film on a silicon wafer
measured by
the AFM is larger as the one found by an X-ray reflectivity
experiment.  The authors thus concluded that the AFM tip
distorts the film and induces a
"bump" in its surface.
At sufficiently high temperatures, even
stronger effects like the formation of a neck between the
adsorbate and the tip have been observed~\cite{kuipers}. Usually,
such a distortion of
adlayers is unaccounted for
while interpretating the experimental data, although its effect
might be not negligible - "condensation" of the adlayer
particles in the vicinity of the tip
would increase the force exerted by the
monolayer on the AFM tip.
Thus the question arises of how to
interprete the AFM measurements
adequately
and how to extract, in a reliable
fashion, the pertinent parameters (say, the Hamaker constants)
in case when some adsorbate is present
on the solid surface.

In this paper we study perturbations of atomic
monolayers on solid substrates induced by {\it immobile} defects
of different types.
The monolayer is described
 using a 2D version of
the Frenkel-Kontorova (FK) model,
{\it i.e.}, we view it as a
2D network of particles connected by harmonic springs
in a spatially periodic potential. Note that the original FK-model
(a harmonic chain in a spatially periodic potential) was
introduced more than sixty years ago in order to describe the
motion of a dislocation in a crystal~\cite{kontorova38}. In the
meantime variants of this model were applied to many different
problems including charge density waves~\cite{Mazo96}, sliding
friction~\cite{Strunz98,Braun97}, ionic
conductors~\cite{pietronero81,Aubry83} and chains of coupled
Josephson junctions~\cite{Mazo96,watanabe96}. A 2D
version
of the FK-model has been introduced by Uhler and Schilling
to study glassy properties of an adsorbed atomic
layer~\cite{uhler97}.

We consider first the case of a surface with a monoatomic step
focusing on two opposite limits: (a) monolayers with strong
intralayer coupling and negligible interactions with the substrate
(smooth structureless surface) and (b) monolayers in which
coupling to the substrate dominates the particle-particle
interactions. As a second example we calculate the perturbation of
a monolayer induced by an immobile AFM tip and demonstrate how it
modifies the force exerted on the tip.

Note that considering the simplified case with an immobile AFM tip
allows us to determine explicitly inhomogeneous density profiles
as well as to elucidate the physics behind this effect. In "real
stuff" experiments the tip, of course, moves and the situation is
fairly more complex since the density profiles are non-stationary.
One expects that for moderate tip velocities the "condensed"
region in the monolayer would travel together with the tip
exerting some frictional force on it. For larger velocities, such
a condensed region would  not have enough time to be formed and
the monolayer should remain homogeneous. One expects hence the
existence of a threshold tip velocity below which the monolayer
has time to reorganize itself leading to an extra force and above
which this effect disappears, a type of force-velocity relation
that is somewhat reminiscent of "solid friction" behavior. A
qualitatively similar behavior has been predicted by Raphael and
de Gennes~\cite{elie} for a system involving a charged particle
moving at a constant speed a small distance above the surface of
an infinitely deep liquid. The situation with a stationary moving
AFM tip will be discussed elsewhere~\cite{heli}.

We constrain ourselves here to the limit of the localized
adsorption~\cite{adamson} and suppose that the adatoms always
remain in close contact to the surface, such that their
defect-induced displacements in the vertical direction
(perpendicular to the surface) are negligibly small. For
simplicity, we assume that the adatoms form a regular
square lattice. Each particle is labeled by two integers, $%
\left( n,m\right) $, with $n,m=0,\pm 1,\pm 2,...$.
For small perturbations of the monolayer the
interaction between a given atom $\left( n,m\right) $ and its four
neighbors $\left( n-1,m\right) $, $\left( n+1,m\right) $, $\left(
n,m-1\right) $ and $\left( n,m+1\right) $
can be represented by Hookean springs that connect
each atom to its neighbors.
The value of the effective spring
constant $K$ follows
from the expansion of the
interaction potential between atoms
near the equilibrium distance
and is typically of the order of a few tenth of
$eV/$\AA$^2$ \cite{Braun97}. In the absence of any external
perturbation the position of atom $\left( n,m\right) $ is given by
the two-dimensional vector ${\bf r}_{nm}=\left(
x_{nm},y_{nm}\right) =\left( bn,bm\right) $ with $b$ being the
equilibrium distance between atoms. In the following we calculate
the defect-induced displacements ${\bf a}_{nm}=\left( \xi
_{nm},\eta _{nm}\right) $ of the adatoms.

First we consider the equilibrium properties of a
monolayer near a
steplike defect (Fig. 1). The
substrate has the height $z=0$ for $%
x>0$ and $z=h$ for $x\leq 0$. We focus on the monolayer on the lower terrace
 ($x>0$). A given small volume element $dV$ of
the substrate is assumed to exert a force ${\bf df}=-adVr^{-\left(
\alpha +1\right) }{\bf \hat{r}}$ on a particle in the monolayer at
a distance $r$ apart; here $a$ is a constant, $\alpha +1$ is an
arbitrary positive number and ${\bf \hat{r}}$ is the unit vector
in the ${\bf r}$-direction. In the absence of a step, $h=0$, the
interaction of any atom
with the substrate is isotropic with respect to rotations around the $Z$%
-axis. Hence, by summing over all forces between a given atom and
the substrate one finds only a force perpendicular to the surface
but no tangential component. In this case the monolayer is
unperturbed. On the other hand, a step of height $h>0$ results in
net forces to the left for atoms to the right of the step.
Consider an atom at ${\bf r}=\left( x,y,z\right) =\left(
D>0,y,0\right) $. The net force follows from integration over the
additional slab of material:
\begin{equation}
f^{\left( step\right) }\left( D\right) =a\int dV\frac{x}{\left(
x^{2}+y^{2}+z^{2}\right) ^{\alpha /2+1}}\simeq -C_{\alpha }\frac{ah}{%
D^{\alpha -1}}  \label{force3}
\end{equation}
with $C_{\alpha }=\sqrt{\pi }\Gamma \left( \left( \alpha +1\right)
/2\right) /\left( \left( \alpha -1\right) \Gamma \left( 1+\alpha
/2\right) \right) $, where $\Gamma(z)$ is the Gamma-function. The
right hand side of Eq.~\ref{force3} holds for $\alpha>1$ and $h\ll
D$. The first condition is needed to insure  that $f^{\left( step\right)}$ remains
finite. The second condition is fulfilled for small step heights
({\it e.g.} monoatomic steps, $h\sim b$).

\begin{figure}
\includegraphics*[width=8cm]{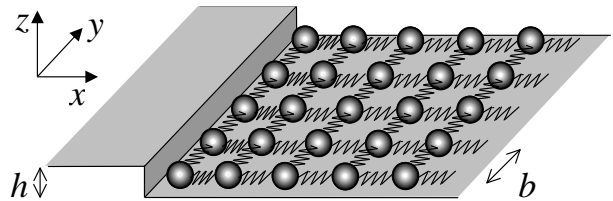}
\caption{Atomic monolayer at a steplike defect of height $h$. The
particles are attracted towards the step and the monolayer is
perturbed accordingly (see text for details)} \label{f.1}
\end{figure}

The positions of the atoms obey the force balance equation $%
x_{n+1m}-2x_{nm}+x_{n-1m}=-K^{-1}f^{\left( step\right) }\left(
x_{nm}\right) $ that can be rewritten using continuous variables
$n$ and $m$ as: ${\partial ^{2}x_{n}}/{\partial
n^{2}}=\left({C_{\alpha }}ah/K{x_{n}^{\alpha -1}} \right)$.
We drop here the $m$-dependence since the problem is apparently
symmetric in the $Y$%
--direction; $x_{n}$ ($n=1,2,...$) denotes the $x$-position of the
$n$-th row of atoms in the monolayer. We denote by $\xi
_{n}=x_{n}-bn$ the displacement of the $n$-th row. Assuming in the
following a weak perturbation of the monolayer, $\xi _{n}\ll b$,
we are led to
\begin{equation}
\frac{\partial ^{2}\xi _{n}}{\partial n^{2}}\simeq \frac{C_{\alpha }ah}{%
Kb^{\alpha -1}}\frac{1}{n^{\alpha -1}},  \label{displ1}
\end{equation}
which yields
\begin{equation}
\xi _{n}\simeq \frac{C_{\alpha }ah}{\alpha \left( \alpha +1\right)
Kb^{\alpha -1}}\frac{1}{n^{\alpha +1}}+B=\frac{l}{n^{\alpha +1}}+B,
\label{an3}
\end{equation}
where $l$ has the dimension of a length. The exact position of the
first row (and therefore the value of the constant $B$) depends on
its specific interaction with the step. One may, for simplicity,
assume $\xi _{1}=0$ and thus $B=-l$. Note that due to the
"coupling" of the different rows the
displacement $\xi _{n}$ increases with $n$ approaching the limiting value $%
B=-l$. For $\alpha =6$ (van der Waals interaction) we find from Eq.~\ref{an3}%
: $\xi _{n}\simeq \left( \pi /480\right) \left( ah/Kb^{5}\right) \left(
n^{-7}-1\right) $.

The  density of adatoms follows $\rho
_{n}=b^{-1}\partial n/\partial x_{n}\simeq b^{-2}\left( 1-b^{-1}\partial \xi
_{n}/\partial n\right) $, which leads to
\begin{equation}
\rho _{n}\simeq b^{-2}\left( 1+\frac{C_{\alpha }ah}{\alpha Kb^{\alpha }}%
\frac{1}{n^{\alpha }}\right),   \label{dens}
\end{equation}
i.e. a
long-range algebraic relaxation to the unperturbed density.
For $\alpha =6$ we find $\rho _{n}\simeq b^{-2}\left( 1+\left( \pi ah\right)
/\left( 96K\left( bn\right) ^{6}\right) \right) $. The density has its
maximal value at the step and decreases with increasing $n$.

Up to now we assumed only an intralayer interaction between atoms
in the monolayer. The role of the supporting substrate was only to
restrict the motion of the atoms in the $XY$--plane. Now we study
the case of {\it strong coupling to the substrate}. We assume next
that each particle in the monolayer is attached to the substrate
via a spring with the spring constant $\tilde{K}$ at the
equilibrium position ${\bf r}_{nm}=\left( x_{nm},y_{nm}\right)
=\left( \tilde{b}n,\tilde{b}m\right) $ ($\tilde{b}$ might be
considered to be the lattice constant of the substrate). We
neglect interactions between neighboring beads, {\it i.e.}, we set
$K=0$. Then the displacement of the particle rows is obtained
directly from the force balance equation $\tilde{K}\xi_{n}\simeq
f^{\left( step\right) }\left( \tilde{b}n\right) $. We find then
\begin{equation}
\rho _{n}\simeq \tilde{b}^{-2}\left( 1-\frac{C_{\alpha }ah}{\left( \alpha
-1\right) \tilde{K}\tilde{b}^{\alpha }}\frac{1}{n^{\alpha -2}}\right)
\label{dens2}
\end{equation}

The perturbation of a monolayer with strong intralayer coupling
(and negligible coupling to the substrate) is fundamentally
different from the case of strong coupling to the substrate. In
the first case -- due to the connectivity of the rows -- the
displacement of each row adds up and the largest displacement, $l$
(in negative $X$-direction), is approached for large $n$-values,
{\it i.e.}, far from the step (cf. Eq.~\ref{an3}). In the latter
case the displacement is directly proportional to the exerted
force which decays algebraically with increasing distance and $\xi
_{n}\rightarrow 0$ for $n\rightarrow \infty $. This is also
reflected in the density profile. The density of the monolayer
with intralayer coupling, Eq.~\ref{dens}, has its largest value
close to the step and decays towards the unperturbed value
$b^{-2}$ for large $n$. On the other hand, monolayers coupled to
the substrate show a slight depletion ($\rho _{n}<\tilde{b}^{-2}$)
close to the step, cf. Eq.~\ref{dens2}. Only very close to the
step (first row of atoms) the density is enhanced accordingly. The
general case with non-vanishing $K$ and $\tilde{K}$ is highly
non-trivial (e.g., in view of possible commensurate-incommensurate
transitions) and is beyond the scope of the present paper.

We study next the perturbation of the monolayer by an AFM-tip
located at height $H$ above the ''central'' atom $n=m=0$. We
assume that the interaction energy between the tip and a particle
at distance $d$ apart is of the form $w\left( d\right)
=-Ad^{-\alpha }$, which yields
\begin{equation}
{\bf f}_{nm}^{\left({\it AFM tip}\right) }=-\frac{A{\bf \hat{r}}_{nm}}{%
d_{nm}^{\alpha +1}},  \label{AFM tip}
\end{equation}
where $d_{nm}=\left| \left( x_{nm},y_{nm},H\right) \right| $ and ${\bf \hat{r}%
}_{nm}$ is the 2D unit vector $\left(
x_{n},y_{n}\right) /\left| \left( x_{nm},y_{nm}\right) \right| $.

The calculation of the elastic force in a monolayer with
intralayer coupling is non-trivial, since the
equilibrium distance of each spring has a non-vanishing value
$b>0$. For
$b=0$ the elastic force is simply given by the Laplacian: ${\bf f}_{nm}^{(spring)}=K%
\left( \partial ^{2}/\partial ^{2}n+\partial ^{2}/\partial ^{2}m\right) {\bf %
r}_{nm}$. For $b>0$ the $X$ and $Y$ direction are coupled in a
non-trivial way. However, the elastic response to small
perturbations ${\bf a}_{nm}=\left( \xi _{nm},\eta _{nm}\right) $
with $\left| {\bf a}_{nm}\right| \ll b$ decouples in the $X$ and
$Y$ directions:
\begin{equation}
{\bf f}_{nm}^{\left( spring\right) }\simeq K\left( \frac{\partial ^{2}\xi
_{nm}}{\partial n^{2}},\frac{\partial ^{2}\eta _{nm}}{\partial m^{2}}\right)
\label{elastic}
\end{equation}
The particle positions follow from the balance between the
tip-monolayer interaction, Eq.~\ref{AFM tip}, and the elastic
force, Eq.~\ref{elastic}, which gives
\begin{equation}
\frac{\partial ^{2}\xi _{nm}}{\partial n^{2}}\simeq \frac{A}{Kb^{\alpha +1}}%
\frac{n}{\left( n^{2}+m^{2}+\gamma ^{2}\right) ^{\alpha /2+1}}
\label{eqmot3}
\end{equation}
where we have introduced the
dimensionless parameter $\gamma =H/b$.
Further on,
%
the displacement $\xi _{nm}$ has to be calculated for each $\alpha $
separately. For $\alpha =6$ (van der Waals forces), for instance, one finds:

\begin{widetext}

\begin{eqnarray}
\xi _{nm}\simeq \frac{-A}{48Kb^{7}}\left( n\frac{3n^{2}+5\left(
m^{2}+\gamma ^{2}\right) }{\left( m^{2}+\gamma ^{2}\right)
^{2}\left( n^{2}+m^{2} +\gamma ^{2}\right) ^{2}} +\arctan \left(
\frac{n}{\left( m^{2}+\gamma ^{2}\right) ^{1/2}}\right)
\frac{3}{\sqrt{m^{2}+\gamma ^{2}}^{5/2}}\right) \label{xn2}
\end{eqnarray}

\end{widetext}

Note that ${\bf a}_{nm}$ is {\it not}
radial-symmetric around $\left( n=0,m=0\right) $ even though it is
induced by a radial-symmetric force, Eq.~\ref{AFM tip}. In fact,
for large $n$ $\xi _{nm}\propto n^{0}$ for $m\equiv 0$ ($X$--direction) and $%
\xi _{nm}\propto n^{-5}$ for $m=n$ (diagonal direction). Such a
non-isotropy appears due to the symmetry of the underlying lattice
(see Fig. 2).

\begin{figure}
\includegraphics*[width=5cm]{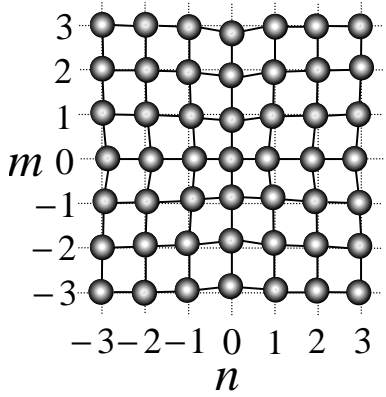} \caption{View from above on a monolayer close to
an AFM tip. The tip is located at the height $H$ above the central
atom ($n=m=0$); see text for details.} \label{f.2}
\end{figure}

For small deformations the density profile of the monolayer is given by
\begin{equation}
\rho _{nm}\simeq \frac{1}{b^{2}}\left( 1-\frac{2A}{\alpha
Kb^{\alpha +2}} \frac{1}{\left( n^{2}+m^{2}+\gamma ^{2}\right)
^{\alpha /2}}\right) \label{rho2d}
\end{equation}
Note that here, however, despite the asymmetry of  ${\bf a}_{nm}$,
the resulting density
profile, Eq.~\ref{rho2d}, recovers the symmetry of the force
exerted by the tip.

We calculate next the force $F$ that the monolayer exerts on the
AFM tip. Due to the symmetry this force is pointed into the
negative $Z$-direction. A particle at $\left( x,y\right) $
contributes to this force by $f_{Z}^{\left({\it AFM tip}\right)
}\left( x,y\right) =-A/H{d^{\alpha +2}}$. The total force $F$ from
the monolayer follows by summing up over all atoms,
$F=F_{0}+\Delta F$ where $F_{0}$ is the force that an unperturbed
monolayer would exert on the AFM tip, $F_{0}=-2\pi A/{\alpha
b^{2}H^{\alpha -1}}$, while  $\Delta F$ denotes the contribution
due to the self-induced perturbation by the monolayer $\Delta
F=-2\pi{A^{2}}/\alpha ^{2}{Kb^{4}H^{2\alpha -1}}$. Note that
$\Delta F/F_{0}=-F_{0}/2 \pi KH$, {\it i.e.}, the induced
additional force is important for soft monolayers (small $K$) and
strong tip-sample interactions.
For the case $\alpha =6$ the two contributions to the force are given by $%
F_{0}\simeq -\left( \pi /3\right) \left( A/b^{2}H^{5}\right) $ and $\Delta
F\simeq -\left( \pi /18\right) \left( A^{2}/Kb^{4}H^{11}\right) $.

We consider now the effect of small surface corrugations of the form $%
U^{\left( surf\right) }\left( x,y\right) =\varepsilon U_{0}\cos \left(
kx\right) \cos \left( ky\right) $ on the positions of the atoms in the
monolayer as well as on the force on the AFM tip. Here $k$ denotes the wave
vector of the periodic substrate and $\varepsilon $ is a small number, $%
\varepsilon \ll 1$. From the potential follows the force that acts
on a particle at $\left( x,y\right) $: $F_{X}^{\left( surf\right)
}\left( x,y\right) =-\partial U^{\left( surf\right)}\left(
x,y\right)/{\partial x}$. We calculate the additional displacement
due to the corrugations using the ansatz $\xi _{nm}=\xi
_{nm}^{\left( 0\right) }+\varepsilon \xi _{nm}^{\left( 1\right)
}$. This leads to $\partial ^{2}\xi _{nm}^{\left( 1\right)
}/\partial n^{2}\simeq -F_{X}^{\left( surf\right) }\left(
bn,bm\right) /K$. Hence
\begin{equation}
\varepsilon \xi _{nm}^{\left( 1\right) }\simeq \frac{\varepsilon U_{0}}{%
b^{2}kK}\sin \left( kbn\right) \cos \left( kbm\right)   \label{xi1}
\end{equation}
As a result of this additional perturbation the force on the AFM tip will be
modified, $F=F^{\left( 0\right) }+\varepsilon F^{\left( 1\right) }$ with $%
F^{\left( 0\right) }=F_{0}+\Delta F$ given above. We give here
explicitly the asymptotic forms of $F^{\left( 1\right) }$ for the case $\alpha =6$:
\begin{equation}
F^{\left( 1\right) } \simeq \left\{
\begin{array}{ll}
\displaystyle \frac{ \displaystyle \pi }{3}\frac{\displaystyle AU_{0}}{\displaystyle b^{4}KH^{5}} & \mbox{for}\;kH\ll 1 \\
\frac{\displaystyle \pi^{1/2}}{\displaystyle 2^{1/4}12}\frac{\displaystyle AU_{0}k^{5/2}}{ \displaystyle b^{4}KH^{5/2}} e^{-\sqrt{2}%
kH} & \mbox{for}\;kH\gg 1
\end{array}
\right.
\label{f13}
\end{equation}
It can be seen from Eq.~\ref{f13} that the contribution from
surface corrugations is ''screened'' when the height $H$ of the
AFM tip exceeds the wavelength $k^{-1}$ of the corrugations. We
dispense with giving a discussion of the case of strong coupling
to the substrate which can be calculated straightforwardly.

\end{document}